\documentclass{article}
\usepackage{graphicx}

\newcommand{\beq}{\begin{equation}}
\newcommand{\eeq}{\end{equation}}

\begin{document}

\title{Did the universe have a beginning?}
\author{Audrey Mithani, Alexander Vilenkin\\
{\small  Institute of Cosmology, Department of Physics and
    Astronomy,}\\ {\small Tufts University, Medford, MA 02155, USA} \\
{\small {\it Email} : vilenkin@cosmos.phy.tufts.edu}
}

\date{}
\maketitle
\begin{abstract} 
We discuss three candidate scenarios which seem to allow the
possibility that the universe could have existed forever with no
initial singularity: eternal inflation, cyclic evolution, and the
emergent universe.  The first two of these scenarios are geodesically
incomplete to the past, and thus cannot describe a universe
without a beginning.  The
third, although it is stable with respect to classical perturbations,
can collapse quantum mechanically, and therefore cannot have an
eternal past.
\end{abstract} 

\section{Introduction} 

One of the most basic questions in cosmology is whether the universe
had a beginning or has simply existed forever.  It was addressed in
the singularity theorems of Penrose and Hawking \cite{HawkingEllis}, with the
conclusion that the initial singularity is not avoidable.  These
theorems rely on the strong energy condition and on certain
assumptions about the global structure of spacetime. 

There are, however, three popular scenarios which circumvent these 
theorems: eternal inflation, a cyclic universe, and an ``emergent'' 
universe which exists for eternity as a static seed before expanding.
Here we shall argue that none of these scenarios can actually be
past-eternal.


Inflation violates the strong energy condition, so the singularity
theorems of Penrose and Hawking do not apply.  Indeed, quantum
fluctuations during inflation violate even the weak energy condition,
so that singularity theorems assuming only the weak energy condition \cite{Borde1994}
do not apply either.  A more general incompleteness theorem was
proved recently \cite{Borde} that does not rely on energy conditions
or Einstein's equations.  Instead, it states simply that past
geodesics are incomplete provided that the expansion rate averaged
along the geodesic is positive: $H_{av} > 0$.  This is a much weaker
condition, and should certainly apply to the past of any inflating
region of spacetime.  Therefore, although inflation may be eternal in
the future, it cannot be extended indefinitely to the past. 

Another possibility could be a universe which cycles through an
infinite series of big bang followed by expansion, contraction into a
crunch that transitions into the next big bang \cite{Steinhardt}.  A
potential problem with such a cyclic universe is that the entropy
must continue to increase through each cycle, leading to a
``thermal death'' of the universe.  This can be avoided if the
volume of the universe increases through each cycle as well, allowing
the ratio $S/V$ to remain finite \cite{Tolman}.  
But if the volume continues to
increase over each cycle, $H_{av} > 0$, meaning that the universe is
past-incomplete. 

We now turn to the emergent universe scenario, which will be our main
focus in this paper.

\section{Emergent universe scenario}

In the emergent universe model, the universe is closed and static in
the asymptotic past (recent work includes \cite{Ellis, Barrow,
  Sergio,Yu,Graham}; for early work on oscillating models see
\cite{Dabrowski}).  Then $H_{av} = 0$ and the incompleteness
theorem \cite{Borde} does not apply.  This universe can be thought of
as a ``cosmic egg'' that exists forever until it breaks open to
produce an expanding 
universe.  In order for the model to be successful, two key features
are necessary.  First, the universe should be stable, so that
quantum fluctuations will not push it to expansion or
contraction.  In addition, it should contain some mechanism to
exit the stationary regime and begin inflation.  One possible
mechanism involves a massless scalar field $\phi$ in a potential
$V(\phi)$ which is flat as $\phi \to -\infty$ but increases towards
positive values of $\phi$.  In the stationary regime the field
``rolls'' from $-\infty$ at a constant speed, ${\dot\phi}=const$, 
but as it reaches the non-flat region of the potential, inflation
begins \cite{Mulryne}.

Graham et al. \cite{Graham} recently proposed a simple emergent model
featuring a closed universe ($k = +1$) with a negative cosmological
constant ($\Lambda < 0 $) and a matter source which obeys $P = w
\rho$, where $-1<w<-1/3$.  
Graham et al. point out that the matter source should not be a perfect fluid, since this would lead to instability from short-wavelength perturbations \cite{Graham}.  One such material that fulfills this requirement is a network of domain walls, which has $w = -2/3$.   Then the energy density is
\beq
\rho(a) = \Lambda + \rho_0 a^{-1}
\label{rho}
\eeq
and the Friedmann equation for the scale factor $a$ has solutions of
the form of a simple harmonic oscillator: 
\beq
a = \omega^{-1} (\gamma-\sqrt{\gamma^2 -1} \cos(\omega t)),
\label{a}
\eeq
where 
\beq
\omega = \sqrt{\frac{8\pi}{3} G|\Lambda|}
\label{omega}
\eeq
and
\beq
\gamma=\sqrt{\frac{2\pi G\rho_0^2}{3|\Lambda|}}.
\label{gamma}
\eeq
In the special case where $\gamma = 1$, the universe is static.
Although this model is stable with respect to classical perturbations,
we will see that there is a quantum instability
\cite{DabrowskiLarsen,MithaniVilenkin}.

\subsection{Quantum mechanical collapse}

We consider the quantum theory for this system in the minisuperspace
where the wave function of the universe $\psi$ depends only on the
scale factor $a$.  In the classical theory, the Hamiltonian is given
by 
\beq
{\cal H} = -\frac{G}{3\pi a}\left( p_a^2 + U(a) \right),
\label{H}
\eeq
where 
\beq
p_a = -\frac{3\pi}{2G}a{\dot a}
\eeq
is the momentum conjugate to $a$ and the potential $U(a)$ is given by
\beq
U(a) = \left(\frac{3\pi}{2G}\right)^2 a^2\left(1-\frac{8\pi G}{3}a^2\rho(a)\right).
\label{U1}
\eeq
With the Hamiltonian constraint ${\cal H} = 0$, enforcing zero
total energy of the universe, we recover the oscillating universe solutions
discussed in \cite{Graham}. 

We quantize the theory by letting the momentum become the differential
operator $p_a \to -i\frac{d}{d a}$ and replacing the Hamiltonian
constraint with the Wheeler-DeWitt equation \cite{DeWitt} 
\beq
{\cal H} \psi = 0.
\eeq
From the Hamiltonian in Eq.~(\ref{H}), the WDW equation becomes
\beq
\left(-\frac{d^2}{da^2}+U(a)\right)\psi(a)=0,
\label{WDW}
\eeq
with the potential from Eqs.~(\ref{rho}) and ~(\ref{U1}).
Note that in quantum theory the form of the potential (see Fig.~1) is no longer
that of a harmonic oscillator. 
\begin{figure}[h]
\begin{center}
\includegraphics[width=9cm]{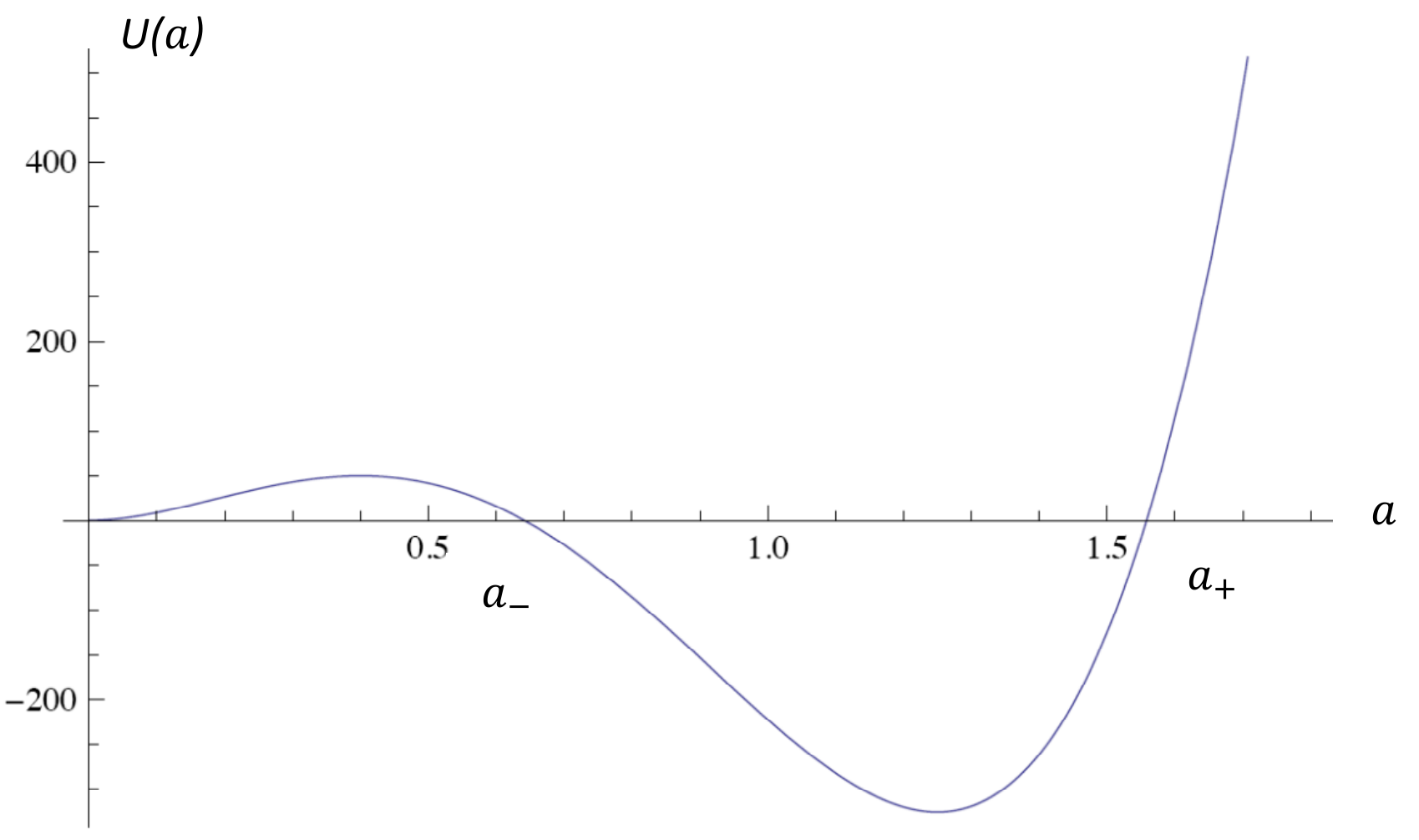}
\caption{The potential $U(a)$ with turning points $a_+$ and $a_-$}
\label{potential}
\end{center}
\end{figure}
Instead, there is an oscillating region between the classical turning
points $a_+$ and $a_-$, which are given by 
\beq
a_{\pm} = \omega^{-1} \left( \gamma \pm \sqrt{\gamma^2 -1} \right),
\eeq
and the universe may tunnel through the classically forbidden region
from $a_-$ to $a=0$.  The semiclassical tunneling probability as the
universe bounces at $a_-$ can be determined
from\footnote{Semiclassical tunneling in oscillating universe models
  has been studied in the early work by Dabrowski and Larsen
  \cite{DabrowskiLarsen}.} 
\beq
{\cal P} \sim e^{-2S_{WKB}}
\eeq
where the tunneling action is
\beq
S_{WKB} = \int_0^{a_-} \sqrt{U(a)}da =  \frac{9M_{P}^4}{16 | \Lambda
  |} \left[ \frac{\gamma^2}{2} + \frac{\gamma}{4}\left( \gamma^2 -1
  \right) \ln\left( \frac{\gamma-1}{\gamma+1} \right) -\frac{1}{3}
  \right]. 
\label{WKB}
\eeq
For a static universe, $\gamma = 1$ and $a_- = a_+ = \omega^{-1}$, 
\beq
S_{WKB} = \frac{3M_{P}^4}{32 | \Lambda |}.
\eeq
Since the tunneling probability is nonzero, the simple harmonic
universe cannot last forever.

\subsection{Solving the WDW equation}

First let us examine the well-known quantum harmonic oscillator.  In
that case, the wave function is a solution to the Schrodinger equation 
\beq
\frac{1}{2}\left(-\frac{d^2}{dx^2} + \omega^2 x^2\right) \psi(x)= E\psi(x).
\eeq
After imposing the boundary conditions $\psi(\pm \infty ) \to 0$, the
solutions represent a discrete set of eigenfunctions, each having
energy eigenvalue $E_n = \left( n+\frac{1}{2} \right)\omega$.
However, in the case of the simple harmonic universe the wave function
is a solution to the WDW equation (\ref{WDW}), which has a fixed
energy eigenvalue $E = 0$ from the Hamiltonian constraint.  From the
form of the potential in Fig.~\ref{potential}, it seems that we must
choose $\psi(\infty) \to 0$, so that the wave function is bounded at
$a\to\infty$. We are then not free to
impose any additional condition at $a = 0$, or the system
will be overdetermined.  The wave function in the under-barrier region
$0<a<a_-$ is generally a superposition of growing and decaying
solutions, and we can expect that the solution that grows towards
$a=0$ will dominate (unless the parameters of the model are
fine-tuned; see \cite{MithaniVilenkin} for more details).

\begin{figure}[h]
\begin{center}
\includegraphics[width=9cm]{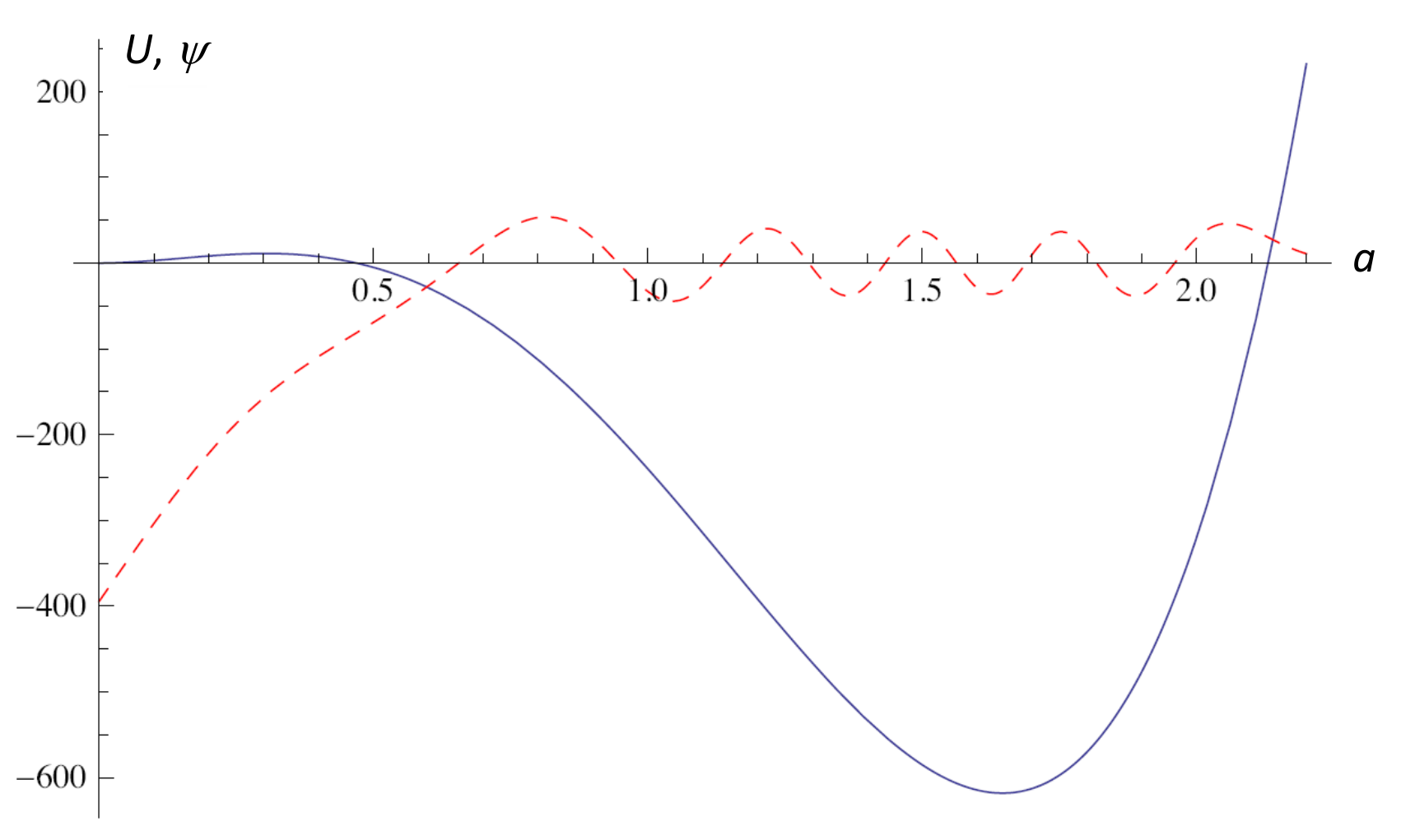}
\caption{Solution of the WDW equation with $|\Lambda| / M_P^4 = .028$
  and $\gamma = 1.3$ (dashed line).  The WDW potential is also
  shown (solid line).} 
\label{oscillatingsolution}
\end{center}
\end{figure}

A numerical solution to the WDW equation is illustrated in
Fig.~\ref{oscillatingsolution}.  It exhibits an oscillatory behavior
between the classical turning points and grows by magnitude towards
$a=0$.  This indicates a nonzero probability of collapse. Similar
behavior is found for the case of $\gamma=1$, corresponding to a
classically static universe.  

One can consider a more general class of models including strings,
domain walls, dust, radiation, etc.,
\beq
\rho (a) = \Lambda + \frac{C_1}{a} + \frac{C_2}{a^2} + \frac{C_2}{a^3}
+ \frac{C_4}{a^4} + \dots. 
\eeq
For positive values of $C_n$, the effect of this is that the potential
develops another classically allowed region at small $a$.  So the
tunneling will now be to that other region, but the qualitative
conclusion about the quantum instability remains unchanged.  Altering
this conclusion would require rather drastic measures.  For example,
one could add a matter component $\rho_n(a) = C_n/a^n$ with $n\geq 6$ and
$C_n<0$.  Then the height of the barrier becomes infinite at $a\to 0$
and the tunneling action is divergent.  Note, however, that such a
negative-energy matter component is likely to introduce quantum
instabilities of its own.

\section{Did the universe have a beginning?}

At this point, it seems that the answer to this question is probably
yes.\footnote{Note that we use the term ``beginning'' as being
  synonimous to past incompleteness.} 
Here we have addressed three scenarios which seemed to offer a way
to avoid a beginning, and have found that none of them can actually
be eternal in the past.  Both eternal inflation and cyclic universe
scenarios have $H_{av} > 0$, which means that they must be
past-geodesically incomplete.  We have also examined a simple 
emergent universe model, and concluded that it
cannot escape quantum collapse. 
Even considering more general emergent universe
models, there do not seem to be any matter sources that admit
solutions that are immune to collapse.

\end{document}